\newfont{\eightbi}{cmbxti10 at 8pt}
\documentstyle[11pt,aaspp4]{article}

\begin{document}

\title{The Three Dimensional Structure of EUV Accretion Regions in 
AM Herculis Stars:\\
       Modeling of EUV Photometric and Spectroscopic Observations}

\author{Martin. M. Sirk\altaffilmark{1}}
\affil{Center for EUV Astrophysics, University of California,
    Berkeley, CA 94720}

\author{Steve. B. Howell\altaffilmark{2}}
\affil{Department of Physics and Astronomy, University of Wyoming,
University Station, Laramie WY 82071}


\altaffiltext{1}{Also with Department of Physics and Astronomy,
San Francisco State University, San Francisco, CA 94132}
\altaffiltext{2}{Guest observer, {\it EUVE} satellite}


\begin{abstract}
We have developed a model of the high-energy accretion region for
magnetic cataclysmic variables and applied it to
{\it Extreme Ultraviolet Explorer} observations of 10
AM Herculis type systems.
The major features of the EUV light curves are well described by the model.
The light curves exhibit a large variety of features such as eclipses
of the accretion region by the secondary star and the accretion stream,
and dips caused by material very close to the accretion region.
While all the observed features of the light curves are highly dependent
on viewing geometry,
none of the light curves are consistent with a flat, circular accretion spot
whose lightcurve would vary solely from projection effects.
The accretion region immediately above the WD surface is a source of
EUV radiation caused by either a vertical extent to the accretion spot, or
Compton scattering off electrons in the accretion column, or,
very likely, both.
Our model yields spot sizes averaging 0.06 R$_{WD}$,
or $f \sim 1 \times 10^{-3}$ the WD surface area, and average spot heights
of 0.023 R$_{WD}$.
Spectra extracted during broad dip phases are softer than spectra
during the out--of--dip phases.
This spectral ratio measurement leads to the conclusion that Compton
scattering, some absorption by a warm absorber, geometric effects,
an asymmetric temperature structure in
the accretion region and an asymmetric density structure of the accretion column
are all important components needed to fully explain the data.
Spectra extracted at phases where the accretion spot is hidden behind the limb
of the WD, but with the accretion column immediately above the spot still visible,
show no evidence of emission features characteristic of a hot plasma.
\end{abstract}



\section{Introduction}

AM Her stars are a class of interacting binaries consisting of a highly
magnetic (typically $10-60$ MG) white dwarf (WD) primary and a red main
sequence
secondary that fills its Roche Lobe. Stellar material flows through the
inner Lagrangian point (L1) and falls toward the primary forming an accretion
stream.
In the absence of a magnetic field, an accretion disk normally forms.
However, in AM Her systems, the strong magnetic field captures the
ionized material and channels it directly toward one or both of the
magnetic poles of the WD primary, forming a hot accretion region.
The interaction of the accretion stream with the magnetic field
circularizes the orbits of both stars and synchronizes the WD rotation
period with the binary orbital period.
When viewed from the rotating frame, the primary, 
secondary, and magnetic
field all appear static.
The only motion is that of the in-falling material, its free-fall time
being about one fourth the orbital period.

Conversion of the kinetic energy of the stream manifests itself
in many forms.
 Electrons in the stream near the accretion region spiral around the
field lines and emit highly polarized cyclotron radiation at optical
and near infrared wavelengths.
X-rays arise in a shock region where the supersonic stream
gives up energy, becomes heated and flows subsonically onto the WD surface.
Extreme ultraviolet (EUV) radiation is emitted from, 
or very close to, the heated WD surface;
the consequence of reprocessed X-rays and the direct mechanical heating
of the WD photosphere by the impacting material.
See Liebert \& Stockman (1985), Cropper (1990), and Warner (1995) for reviews on AM Her
systems.


The field strengths of the white dwarfs in magnetic CVs are strong
enough to focus accreted material onto their surface in a relatively
small area ($f \sim 1 \times 10^{-3}$ of the WD surface) called the accretion region. 
Observational evidence
for this situation comes from many sources, all of which provide convincing
proof for the white dwarf spin to be locked with the binary orbit
and for the compactness of the accretion region.
The accretion onto a magnetized white dwarf is nearly a radial in-fall
situation, making the problem almost one-dimensional in nature. 
Radial accretion onto a WD surface was first considered to explain X-ray
emission in Sco X-1 (Cameron \& Mock, 1967). 

The general idea of magnetically focused accretion 
is as follows. Ionized material
leaving the L1 point of the secondary is in free fall until at some
point, called the coupling region, the kinetic energy associated with
the angular momentum is overcome
by the magnetic energy of the primary field, and the material gets funneled
along the field lines to the WD surface.
The flow
can initially be assumed to be uniform from the secondary all the way to
the primary star. We will see that this is likely {\it not} to be the
case but will serve as a working model for now. A strong shock is
encountered by the accreting stream just above the surface where it
decelerates by a factor of 3-4, converting much of its infall energy to
short wavelength radiation.
Post--shock material moves at subsonic velocities and settles onto
the WD surface. Temperatures in the post--shock region are near kT$_{TB}~\sim$
10-25 keV and kT$_{BB}~\sim$20-40 eV, with 
emission peaking in the X-ray and EUV regions. This short wavelength 
radiation leaves the
accretion column easily, with about half escaping into space and half
going towards the WD photosphere around the accretion region. 
The accretion area 
near the WD surface becomes heated to a few times $10^5$~K, giving
rise to EUV emission. 

The accretion region was originally modeled as a circular region
(often referred to as an accretion spot)
lying flat on the WD surface (Lamb and Masters, 1979).
These authors also calculated
the overall emitted spectrum as a sum of cyclotron emission,
bremsstrahlung, and reprocessed black-body emission from the heated
surface. Observations of AM Hers at high energies (X-rays), revealed a
fascinating array of complexities in the accretion regions. By the late
1980's, new ideas for the size and shape of the accretion regions were
emerging. Foremost among these was the work of
Wickramasinghe and Meggitt (1985), Wickramasinghe, Ferrario \& Bailey (1989),
and Ferrario, Wickramasinghe \& Tuohy, (1989). 
Their view gave the accretion spot an arc-like
shape of length $\sim \frac{1}{4}$ R$_{WD}$, instead of a circular
profile, regions of high and low density mass flow, and the spot had a small
height ($\sim \frac{1}{10}$ R$_{WD}$) above the surface.
Surrounding the entire spot was an extended corona-like halo of hot plasma .
 
The early models of a homogeneous flow with cylindrical symmetry
in the accretion stream made the
prediction that the emitted luminosities at high energy will be such
that the hard X-ray emission will be greater than about 2 times the soft
X-ray (EUV) emission. 
Observations in the EUV spectral region, particularly in
recent years by the R\"ontgen Satellite ({\it ROSAT}) and the
 Extreme Ultraviolet Explorer ({\it EUVE}), have again caused
accretion region models to be re-evaluated (eg. Ramsay et. al. 1994,
Beuermann \& Schwope 1994).
 These authors and others, have shown that
soft X-ray excesses exist in most of the AM Herculis systems,
unexplainable in the older models.

Discussions of the ``soft X-ray problem'' are given in Cropper
(1990), King (1995), and Beuermann \& Burwitz (1995) and references cited
therein. New ideas including inhomogeneous flows, blobby accretion,
buried accretion regions, and cyclotron cooling of the post--shock region
have been pursued in an attempt to understand the observations.
Blobby accretion was first discussed in Kuijpers \&
Pringle (1982) and followed up on by others. Litchfield (1990),
allowed the blobs to bury themselves deep into the WD
atmosphere before giving up their store of energy, causing increased
local heating and giving rise to an ``excess'' of EUV photons.
Litchfield also calculated a shape and size for the accretion region, 
and the lightcurve
expected for 100~\AA\ observations (see \S 7.2). 

King (1995) has suggested that one can match the observations via an
accretion region consisting of various sites of bombardment within the
total effective radiating area, the latter being the area in which the
blobs usually land. Each of the sites may become a depressed accretion
region, buried beneath the WD surface by several atmospheric scale
heights. 

All of the models discussed above generally agree that the accretion region is
relatively small and its ``size'' is wavelength dependent. The hard
X-rays come from the central concentration, presumably located
near the stream shock. Surrounding this central concentration,
there is an extended region of EUV emission 
($\sim 1000$ km in extent), heated from
above by X-rays from the shock and from below by thermal energy released
by the impacting blobs. (see Cropper;1990, Stockman et. al.; 1994, \& 
Schwope; 1995).


Disentangling the various radiations from their points of origin is
difficult, especially in the optical where the photospheres of both
stars as well as the accretion stream all contribute flux.
Observations in the EUV (65\AA\  to 180\AA) 
are particularly well suited to probing the
accretion region since the source of radiation is often confined to a very
small region on, or near, the WD surface. Furthermore, many AM Her systems
are intrinsically bright around 100~\AA .
The peak energy release
of the extended spot, and the pre- and post-shock regions of the stream
are all sources of EUV emission.

The {\it EUVE} satellite
has performed a number of pointed photometric and spectroscopic
observations of AM Herculis stars (cf., Craig et. al., 1997),
and in this paper we present results for many of these stars.
Using the high quality, high time-resolution EUV data, we have developed a new
model
for the accretion geometries of AM Her systems. This model, discussed
in \S 4, provides quantitative information on the size,
shape, and variability of the accretion region. In addition, our model provides
the first direct evidence of extended EUV emission above the WD surface,
originating in either an extended halo, or the lower (pre--shock) portion 
of the stream itself.


\section{Geometric Variability of Accretion Regions}

Multi-wavelength light curves of AM Her stars show strong
modulation arising from both geometrical and physical effects.
The intrinsic amount of radiation from AM Her stars varies on time scales
ranging from seconds to years and is mainly controlled by the mass transfer rate.
During one binary orbit, the viewing geometry to the accretion region changes
considerably.
The angle subtended between the observer and the accretion spot normal
vector varies as a function of the system inclination $\iota$,
the spot colatitude $\beta$ (measured from 0\arcdeg\ to 180\arcdeg\
from the rotational pole tilted towards the observer),
and the rotation phase $\phi_b$ of the synchronously locked binary (see Fig. 1).
For a small ($r \ll$ R$_{WD}$) flat accretion region confined to the WD surface,
the observed modulation is purely sinusoidal as a function of phase
and reaches maximum brightness when the viewing angle is at a minimum
(that is, when the spot normal is pointed most directly towards the observer).
This occurs when the accretion spot crosses the central
meridian of the WD and presents its largest cross section.
In systems where $(\iota + \beta) > 90\arcdeg$ a flat accretion spot rotates
behind the limb of the WD surface and is completely occluded
for some portion of the binary orbit.
In Figure 2 we show synthetic light curves for a small accretion region
illustrating the sinusoidal
modulation caused by changes in viewing geometry for 
different system inclinations
and accretion spot colatitudes.
We discuss, in more detail, the model light curves of Figure 2
in \S 4.1.

Highly inclined systems ($\iota \approx > 73\arcdeg$)
also show eclipses of the WD primary by the secondary dwarf star.
If the accretion region appears in the upper hemisphere ($\beta < 90\arcdeg$) of
the WD, then the accretion stream, when it is still far from the WD surface
(the far-field stream) {\it must} cross the line of sight
to the accretion region at some point during the orbit
and may result in a dip
in the lightcurve (Mason, 1985, King and Williams, 1985).
These far--field dips often saturate to zero flux mimicking a stellar eclipse.
Finally, the accretion stream in the immediate vicinity of the
accretion region (referred to as the pre--shock or near--field accretion column)
may also absorb or scatter EUV radiation out of the line of sight
producing dips in the lightcurve (Imamura \& Durisen, 1983, hereafter ID83).
These near--field accretion dips tend to be broad in phase and 
generally do not saturate to zero flux.
The amount of attenuation is strongly dependent on the viewing geometry
through the column to the spot, being greatest when one looks parallel
to the column, and least when viewing perpendicular to the column.

We present in Figure 3, the 1995 January {\it EUVE} lightcurve for
the eclipsing system UZ Fornacis to exemplify the features discussed
in this paper.
Clearly, the light curve of UZ For is more complicated than the simple
sinusoidal variations predicted by a basic model in Figure 2.
Identifying the causes of the observed modulations and features in the
EUV light curves by developing a general model
of the accretion region applicable to any AM Her star is 
the goal of our analysis presented here.

Analysis of {\it Extreme Ultraviolet Explorer} ({\it EUVE};
Bowyer \& Malina 1991)
photometry of UZ For
(Warren, Sirk, \& Vallerga 1995, hereafter WSV)
showed a small ($<$ 0.23 R$_{WD}$) EUV accretion spot that is 
raised a few percent above the WD surface.
Furthermore, the symmetric eclipse profiles of the spot precluded
any large scale longitudinal structure
(such as an auroral arc with a brightness gradient).
Vennes et. al., (1995) also determined a height of a few percent of the R$_{WD}$
for the accretion spot of VV Pup based on model atmosphere calculations.
Vertical extension of the accretion region profoundly affects the EUV light
curves (cf., Sirk \& Howell 1996), both by lengthening the duration of the bright phase (when the
spot is not behind the WD limb), and steepening the rise and fall phases
(when the spot rotates into and out of view from behind the WD limb).
To understand the physical processes of AM Her systems one requires an accurate
knowledge of the system geometry, namely the orbital inclination ($\iota$)
and the colatitude of the accretion spot ($\beta$).
Determination of these parameters is often done via optical light curve and
polarimetry observations.
The assumption of a flat accretion spot when analyzing these types of data
may result in incorrect values of
$\iota$ and $\beta$.
Our initial analysis modeled the EUV accretion regions, allowing
the inclination, spot colatitude, spot size, and spot height to all
be free parameters.

\section{Data and Light Curves}

The observational data for the ten systems analyzed here consists of our 
own {\it EUVE} guest observer data
(Howell et. al., 1995, Sirk et. al., 1998),
data from the {\it EUVE\/} public archive,
and data from the {\it EUVE\/} Right Angle Program (McDonald et. al., 1994).
All of the photometric data used was obtained in the Lexan/Boron passbands
of the deep survey and scanner telescopes
($\geq$ 10\% of peak transmission from 67 to 178 \AA, peak at 91 \AA; 
Sirk et. al., 1997) and
the short wavelength spectrometer
($\geq$ 10\% of peak transmission from 65 to 178 \AA, peak at 100 \AA; 
Boyd et. al., 1994).
The deep survey and three spectrometer channels of {\it EUVE} are co--aligned;
photometry and spectroscopy for a given target are obtained simultaneously.
An observation log of the ten systems studied is presented in Table 1.
The source photons are counted in a circle usually of 1 arcmin in radius,
and the background estimated in a concentric annulus with typical inner
and outer radii of 3 and 11 arcmin. 
To preserve the high temporal resolution of
{\it EUVE}, the arrival time of each photon is first folded by the 
appropriate binary period,
then a histogram is constructed at some chosen phase bin interval
(ranging from 3 s for the bright sources to 12 s for the faintest source).
Heliocentric corrections  are applied to every photon arrival time
before folding onto the binary periods.
Finally, the exposure time (corrected for detector deadtime)
is calculated for each phase interval
(see Hurwitz et. al., 1997 for details).

The EUV light curves of the AM Herculis stars fall into two distinct categories:
the first is where the accretion
region rotates behind the limb of the WD for some portion of the phase
(UZ For, HU Aqr, VV Pup, RE1844-74, RE1149+28, and AM Her, $\beta + \iota > 90\arcdeg$),
and the second is where the accretion region remains visible at all phases
(EF Eri, AN UMa, and V834 Cen, $\beta + \iota < 90\arcdeg$).
Qualitatively, the first category of light curves resemble the 
positive portion of a cosine curve, and generally show one or two dip features.
The second category (which should resemble an entire cosine cycle)
are dominated by three or four dip features each.
We present our sample of EUV 
light curves for six self eclipsing AM Her stars in 
the first eight panels of Figure 4.
The non-self eclipsing systems are shown in the 
remaining four panels of Figure 4.
The plotted symbols (open and filled circles)
represent the observed countrate at a given binary orbital phase
and the solid and dashed lines represent model fits: 
both and are discussed in \S 4.

One object, QS Tel (Rosen et. al., 1996), shows evidence for accretion onto 
only one
pole during one epoch, and onto both magnetic poles at other times (Fig.4).
Since its inclination is not yet known, we can not classify it uniquely into 
either EUV light curve family nor fit it with our model. 
Two of the systems are high inclination (UZ For \& HU Aqr) 
and have the WD and accretion
region totally eclipsed by the secondary star.
Eclipse timings at all wavelengths provide stringent limits on the system
inclination, and constrains the horizontal size
of the accretion region.
Analysis of the self eclipse of the accretion region by the WD limb in 
the first category of light curves provides a direct measure of the spot height and
its vertical brightness profile, as well as constraints on
the system inclination and spot colatitude.
In the second category of light curves
these same parameters are difficult to determine as they are poorly constrained.
Thus, we have concentrated our initial modeling efforts on the first category
of systems.

Since the total on--source integration times were
significantly longer than the binary orbital
periods (by factors of 5 to 35),
all phases are sampled multiple times (although not necessarily equally).
Therefore, the phase folded light curves 
represent time averages, not instantaneous
states.
The scatter in the countrates for a single binary orbit are 
much larger than that
of the mean light curves shown in Figure 4 (eg., see Howell et al.,
1995).
Analysis of the short time scale variations will be 
the subject of a future paper.

\section{Light Curve Modeling} 

	\subsection{Geometric Accretion Spot Models}

All of the AM Her stars presented in Figure 4 that 
undergo self eclipse,
when the accretion spot rotates behind the limb of the WD,
show a symmetric EUV rise and fall phase when reflected
about the EUV mid--phase point.
The symmetry is so remarkable, in fact, that we can define 
the phase midway between the
rise and fall phase of the light curves,
when the accretion region lies on the central meridian of the WD
pointing most directly towards the observer,
as phase 0 throughout this analysis.
Enlargements of the light curves of the rapid rise and fall phases are
presented in
Figure 5 with the fall phase mirror reflected about
the EUV mid--phase.
In all six systems, the early rise phase and the late fall phase show almost
a perfect match,
which indicates that the accretion regions cannot possess any large 
scale horizontal asymmetries.

The two eclipsing systems 
(UZ For, WSV; and HU Aqr,
Sirk et. al., 1998, Schwope et. al., 1998)
show very rapid ingress and egress eclipse times in the EUV of $\sim$ 2 s
each. This short time scale constrains the size of the accretion regions 
to 16\% and 23\%  of the R$_{WD}$,
respectively, and restricts the region of EUV emission to within
$\approx 0.15$ R$_{WD}$ of the WD surface.
This strong evidence for a small, symmetric, vertically extended accretion spot
in the two eclipsing systems,
and their similarity to other non--eclipsing AM Her systems, motivated us
to model the EUV accretion regions in all ten AM Her stars 
with simple geometrical models.

Three symmetric, uniformly bright accretion spot models were constructed;
A flat circular spot, a cone with circular base, and a raised mound
(sector of rotation).
In each model, the brightness is directly proportional to the cross--sectional
area visible at a given phase (ie., in this first stage, we assume the column is
optically thin and the spot exhibits no limb darkening.
We explore the effects of an optically thick column in \S 4.3).
The synthetic light curves from a small, flat spot closely resemble the
positive portion of a cosine curve,
and thus, completely fail to account for the steep rise and fall
phases of the EUV light curves (as well as any dips).
The conical spot model also failed to reproduce the steepest portions
of the light curves.
A raised mound, however, fit the rise and fall phases very nicely.
When seen face on, the mound model is indistinguishable from a flat spot.
However, when viewed edge on, a flat spot disappears, whereas a mound still
shows appreciable area and continues to do so as the footprint of the spot
rotates behind the WD limb.
We explored a full range of spot sizes, spot heights and colatitudes and present
a typical example in
Figure 2 showing synthetic light curves for a flat circular spot
and a raised mound at four different binary inclinations
$\iota =$\ (80, 60, 40 and 20\arcdeg) and a range of spot colatitudes spanning
both hemispheres.
In this example the spot radius is 0.06 R$_{WD}$, and
for the mound model the height is 0.03 R$_{WD}$.
We also created asymmetric spots by placing two spots of identical radius,
but one twice as bright as the other, side-by-side along the direction
of rotation.
Noticeable departures from symmetry in the synthetic light curves
occurred when the long axis of the
composite spot model exceeded 0.2 R$_{WD}$.
This type of large scale asymmetric accretion region seems to be completely ruled
out by the observations.

\subsection{Application of the Geometric Accretion Spot Model} 

The geometric spot model for a symmetric accretion region 
(flat or raised) produces
symmetric light curves.
The six systems shown in Figure 4 that 
undergo self eclipse where
the accretion spot rotates behind the limb of the WD,
show a symmetric EUV rise and fall phase when reflected
about the EUV mid point.
This fact alone indicates that, regardless of whether the accretion spots are
flat or show vertical extent,
they all must be small ($r_{spot} < 0.2$ R$_{WD}$) in longitudinal extent, or
have a uniform brightness profile in the longitudinal direction,
since the observed symmetry would not be produced by a large accretion region
extended in magnetic longitude with a non-uniform brightness profile.
The steep, nearly linear rise and fall phases seen in the self eclipsing
systems (Figs. 3--5) are inconsistent with 
the gently curving cosine behavior expected
from projection effects of a flat spot (Fig. 2)
and indicate a vertical extent of the EUV emission above the accretion region.

We applied our geometric model first to UZ For since its inclination
and spot colatitude were
accurately determined from optical eclipse photometry by
Bailey \& Cropper (1991), and the EUV spot size is known from eclipse
ingress and egress timings (WSV).
To test the utility of the model, we allowed the inclination and the spot
colatitude 
to be free parameters, even though they were already known.
The central parts of each light curve are subject to eclipses by
the secondary star, the far--field accretion stream, and
show other dip features  probably caused by
material very close to the accretion spot itself.
We initially excluded these regions from the model fits
since it is impossible to reproduce them from projection effects alone
(we explicitly address these features in \S 4.3).
We fit the filled circles (Fig. 4) using our raised mound model and 
an iterative, non--linear least squares fitting routine. The algorithm requires
initial starting values for each parameter.
To ensure convergence at the global minimum, and not a local minimum of
the parameter space,
the fit is iterated repeatedly with different starting values.
Fitting our raised mound model to UZ For,
we find a spot radius and spot height of .06 and .03
R$_{WD}$, respectively, and an inclination and colatitude of 80\arcdeg\  and
136\arcdeg.
These latter two values are in excellent agreement with those determined
by Bailey \& Cropper (1991).
The same model was then applied to HU Aqr, VV Pup, AM Her, 
RE1149+28 and RE1844--74.
Our fits are shown as dashed lines in Figure 4.
For the two eclipsing systems, UZ For and HU Aqr, where the absolute phase zero
is known (defined as the center of the eclipse of the WD by the secondary),
the accretion spot longitude ($\psi$) is obtained directly as the
difference in phase between the center of the eclipse and EUV mid--phase zero.
Our derived system and accretion region parameters are listed in Table 2.


Errors in the fit parameters
are dominated by the assumptions in the model
(i.e., circular symmetry and uniform brightness)
and subjective choices made
in the construction of the light curves, not by the Poisson error in the
actual data themselves.
Phase zero for each system is determined directly from the
light curves by mirror reflecting the fall phase onto the rise phase until
they overlap. This process is done visually and is thus somewhat subjective.
The choice of binary phase bin size 
(mainly dependent on EUV countrate)
also affects the fits slightly.
The subjective error is estimated
by varying the bin size and the phase zero point,
and comparing the resultant fits.
Stable solutions were found for all systems and typical errors for each
parameter determined.
The binary inclination has the greatest effect on the lightcurve models and is
therefore strongly constrained to within 5$\arcdeg$. 
The spot colatitudes are stable to within 7-15$\arcdeg$,
and the spot height and radius to within 25\% and 35\%, respectively.

In Table 3 we compare our system geometry results to those of others
using different methods.
The modeling of optical and near IR polarization data has become quite sophisticated,
providing not just system inclination and accretion spot colatitude, but
also the size and shape of the accretion region
(cf. Wickramasinghe et. al., 1991, Ramsay et. al. 1996, Potter et. al., 1998).
Analysis of spectrophotometric data by the means of Doppler tomography
yields the location of origin of optical line emission,
confining it unequivocally to (a) the secondary star, (b) the
ballistic portion of the accretion stream, or (c) the magnetically
funneled part of the stream (Schwope et. al., 1997).
The deconvolution of the various line velocities imposes
limits on the system inclination and mass ratio.
While the agreement between the values of system parameters seen
in Table 3 is excellent, it must be emphasised that the location of the source
of radiation employed in each method is quite different.
The EUV radiation is confined to a small region very near the WD surface,
whereas the source of optical radiation very likely extends 
much higher up the stream
and contains contributions from other system components.
Thus, any model that aims to accurately predict the system inclination, and the
size, shape, and location of the accretion region must incorporate
a three dimensional source of radiation.
Assuming that the optical radiation emanates from the flat footprint at the base
of the accretion column (as is commonly done in polarimetry modeling) introduces
systematic errors in the determination of the system inclination and spot
colatitude (M. Cropper, 1998, private communication).
Wu \& Wickramasinghe (1992) have made progress in modeling the optical
emission region as a three dimensional structure.
Modeling the EUV accretion region as a raised spot of emission provides
an independent means of estimating the system geometry.

For the six systems where the accretion spot undergoes self eclipse,
the symmetric, raised spot model accounts for the early and late EUV phase
(ie. where the accretion spot is seen mainly edge--on near the WD limb)
very well.
The geometric raised mound model showed us that the vertical extent
of the accretion region is the primary cause of the symmetry in the
early and late portions of the EUV light curves.
Longitudinal brightness gradients, as we have seen, 
have virtually no effect on the light curves
at any phase for raised accretion regions whose longitudinal extent is less
than 0.2 R$_{WD}$.
Hence, any system that shows symmetric rise and fall phases must either have
a small accretion region, or a uniform brightness profile in longitude.


A careful study of the residuals of all the fits to the category 1 systems
showed a systematic and symmetric
underestimate of the flux at the very earliest rise phase, and very latest 
fall phase (cf., Figs. 3 and 4).
The region of residual flux typically spans 10\% of the phase for each star.
Rather than invoke a horizontal brightness profile
of the accretion region that symmetrically leads and follows the
spot center by 5\% in phase on each side,
we attribute the residual flux ($\sim$ 1\% of the total flux)
to EUV radiation emanating from the
accretion region (column) immediately above the accretion spot proper.
Our data are consistent with the brightness of the column decreasing
exponentially with distance from the WD surface.
Thus, we need add only two additional parameters (a brightness scale factor,
and an exponential decay constant) to our raised mound 
model to adequately reproduce the
residual portion of the light curves.
The maximum height of this extended EUV luminous accretion column
may be measured directly from the light curves once $\iota$ and $\beta$
are known.
Figure 6 compares the fits of a flat circular spot,
a circular raised mound, and a circular raised
mound with luminous accretion column to the fall phase of UZ For.
The flat spot fit shows large residuals and places the accretion spot
$130\arcdeg$ away (in the opposite hemisphere) from its known location.
The raised-mound fit does well in the steep fall phase
but systematically underestimates the late fall phase.
The final panel models the accretion column as a source of EUV flux
that decreases exponentially with height above the WD surface.
All points in the rise and fall phases are fit by this model.

In Figure 7 we plot our derived accretion column intensity as a function
of distance above the WD surface for three systems, normalized to the peak
intensity measured for each system.
The source of this vertically extended EUV emission is not well
understood,
but later in this paper, we discuss the likely possibility 
that the excess EUV flux at these phases is from 
Compton scattering by electrons in
the near-field accretion column. Emission (lines) from a
corona-like halo surrounding the accretion region \footnote {Evidence
for such coronal emission is seen in the DQ Herculis star EX Hya
(Hurwitz et. al., 1997) and possibly in PQ Gem (Howell et. al., 1997)}
are predicted by theory, but as we will see, are unobserved in AM
Hers.

	\subsection{Near-- and Far--Field Accretion Column Effects} 

A quick look at Figure 4
shows that the EUV light curves in general are anything
but symmetric.
As the accretion spot rotates towards the central meridian and is viewed
mainly face-on, the light curves show a variety of dips which cannot
be accounted for by any simple geometric model.
Every system shows at least one dip, and some two or more.
For systems where $(\beta-\iota) < 0\arcdeg$ the far--field accretion stream is
guaranteed to
cross the line of sight to the accretion spot possibly causing a dip.
For highly inclined systems ($\iota \approx > 73\arcdeg$) the stream may
still intersect the line of sight even if $(\beta-\iota) > 0\arcdeg$
due to the finite
width of the stream (WSV).
This is indeed the case for UZ For and VV Pup;
both systems show deep stream dips (labeled as such in Fig. 4),
high binary inclination, and accretion
spots in the lower hemisphere ($\beta > 90\arcdeg$).
Because the in-falling material retains the angular momentum of the 
secondary star, the accretion stream must lead 
the line connecting the star centers in the direction of binary rotation.
Thus, any observed eclipses of the
accretion spot by the far--field stream must occur before absolute phase zero.
Since the accretion spot in AM Her stars generally leads absolute phase zero by
roughly 45\arcdeg\ (Cropper, 1990),
the far-field accretion stream dip will occur near the
middle of the EUV bright phase 
(phase zero throughout this analysis).
The observed stream dips are narrow in phase and imply that the absorbing
material is far from the WD surface.

In addition, broader dips are seen in every system, occurring for the most part
at earlier phases than the far-field stream dips
(labeled ``Dip'', and ``Stream Dip'', respectively, in Fig. 4). 
A firm conclusion from our model is that {\it none} of the observed dips
can be reproduced merely from differences in viewing geometry.
As the accretion spot emerges from the WD limb, its projected area
(and hence its brightness) can only {\it increase} until it crosses
the central WD meridian, regardless of the spot's shape or horizontal
brightness profile (see Fig. 2).
The observed width of the broad dips suggests that light is being removed from
the line of sight by material close to the point of origin.
Thus, we hypothesize that absorption or scattering by the near--field 
accretion column itself, in the immediate vicinity
of the accretion spot, is the cause of these features. 
As the WD rotates, the path length (both geometric and optical) 
along the line of sight
through the near--field column to the accretion spot varies continuously
and considerably.

To account for the broad dip, 
we construct an absorption model that assumes a cylindrical
accretion column of uniform density
incident upon the WD surface at some angle with respect to the spot normal,
and with a radius equal to the spot radius.
Two angles are required to uniquely specify the direction angle of incidence
of the column relative to the spot normal.
The first is the angle of incidence of the column tilted in the direction of latitude.
The second angle is the angle of incidence of the column tilted in the direction of
rotation (longitude) and is estimated directly
from the light curves as the
difference in phase between the center of the broad dip
 and the EUV mid--phase (phase zero).
This latter angle is input to the model as a fixed parameter.
Figure 8 illustrates the geometry of this model
which requires just two free parameters;
the angle of incidence of the column tilted in the direction of 
latitude, $\alpha$,
and an absorption coefficient $\tau$ where the transmission through
the column obeys
$$
T=e^{-\tau l}\ .
$$

The effective transmission is calculated by integrating over all possible
path lengths $l$ through the cylindrical column to the spot
for a given viewing direction.
The path length $l$ is determined analytically from the rotation phase $\phi_b$,
the inclination $\iota$, the spot colatitude $\beta$, and the angles
of incidence of the column in latitude $\alpha$ and longitude,
and the column (spot) radius $r$.
By varying the parameters $\tau$ and $\alpha$,
symmetric dips of nearly any width and depth are possible.
Since the values of inclination, spot colatitude, and spot height have already
been determined by the geometrical model, 
we hold them constant as fixed
parameters.

Using this straightforward absorption idea, in connection with our
previous spot model, five systems (excluding HU Aqr since
it showed no obvious broad dip in the May 1996 observation\footnote { 
However, recent EUV observations of HU Aqr, obtained in May 1997, 
showed a pronounced
dip quite similar to that seen in UZ For and will be addressed in Sirk et.
al., 1998}) were modeled well.
We again exclude from our fits data points at phases where 
the narrow dip by the accretion stream far from the WD (and the stellar
eclipse seen in UZ For) dominate.
The ratio of the total absorber model flux to the total un--absorbed model flux
(ie. the ratio of the areas under the solid and dotted line fits in Figure
4),
integrated over the phase interval where the accretion region is not hidden
behind the white dwarf,
and the derived values of the absorption coefficient $\tau$ are
presented in the final two columns of Table 2.
The optical depths determined by modeling the broad dips range from 0.1 to 2.
The tilt angles ($\alpha$) of the accretion column with respect to the spot
normal ranged from 4\arcdeg\ to 51\arcdeg.
They are not listed in Table 2 since we do not believe that accretion columns
generally tilt more than $\sim$10\arcdeg\ (eg., Meggitt \& Wickramasinghe, 1989).
However, our model clearly requires large tilt values (in both the longitudinal
and latitudinal directions)
and we discuss the significance of these tilts in \S 7.2.

Our cylindrically symmetric column absorber model accounts for the gross
features of the
broad dips quite well as shown by the solid lines in Figure 4.
We point out, however, that where the light curves in Figure 4 
show sharp transitions
(e.g. the right angle kinks visible at the end of the rise phase in both 
UZ For and VV Pup), 
the residuals from our absorber model fits
indicate that the true situation must be more complex 
than can be accounted for from a cylindrically symmetric,
uniformly dense accretion column.
This is no surprise as the accretion columns are likely to be clumpy,
asymmetric, and have structure to them (cf., Cropper 1990,
Wu \& Wickramasinghe, 1992).
By themselves, the EUV light curves cannot constrain in detail
a more sophisticated
model with additional structure 
but we discuss likely possibilities in \S 7 and are currently developing
a more realistic near--field stream model.

For the second category of light curves (where the spot is visible at
all phases),
an attempt was made to interpret the light curves of
V834 Cen, AN UMa, QS Tel, and EF Eri in terms
of our absorber plus spot model in spite of the fact that
the inclination, spot colatitude, spot size and height are not
directly determinable from the light curves.
We performed a grid search in parameter space,
the mean starting values for
the inclination and colatitude for each system are taken from Cropper (1990),
and the mean spot radius
and height used were an average value found from the first category of systems.
Stable solutions were found for each star, but deciding where the EUV mid-phase
lies, and distinguishing between the far--field stream dips and the near--field
accretion column broad dips,
among the many visible dip features is highly subjective.
Our absorber model (uniform density, symmetric spot and column) 
will only account for
a single dip, whereas these systems show what appear to be multiple
structures.
We conclude that the numerous features in these light curves are indicative of
additional structure in the accretion stream, column,  
and/or in the accretion region
such as multiple column filaments and multiple EUV emission sites
(cf., Harrop-Allin et. al., 1998).

\section{Phase-Resolved EUV Spectra}

Can the EUV spectra tell us anything about the nature of the observed 
photometric behavior seen in the AM Her systems?
Using the light curves as a guide, we extracted phase resolved spectra for
the three bright systems (UZ For, VV Pup, \& AM Her) 
in the manner described in
Hurwitz et. al., (1997).
We selected four distinct phase regions:
the earliest rise and fall phases where the spot is still fully obscured
behind the WD limb with only the flux from the column above the spot visible,
the broad dip region, a bright region at late phase where the
light curves show no dips, and
for UZ For and VV Pup a spectrum during the far--field stream dip phase.
These phase regions are delineated in Figures 3 and 4.

The EUV spectra averaged over the bright phase 
for AM Her stars typically show an
(interstellar) 
absorbed blackbody continuum without any emission lines and possibly
weak absorption features (Paerels et. al., 1996,
Vennes et. al., 1995). This is in striking contrast to the EUV spectrum of the
intermediate polar star EX Hya which is dominated by iron emission lines
characteristic of a plasma with temperature $10^7$ K (Hurwitz et. al., 1997).
The EUV 
spectra from the accretion region near the WD surface of AM Herculis stars
are inconsistent with unperturbed  emission from a hot, low density plasma.

However, what about the portion of the accretion
column above the spot that is visible during the earliest rise and latest fall
phases when the accretion spot footprint is hidden behind the WD limb? 
Hydrodynamic calculations of accretion onto a white dwarf by
Woelk \& Beuermann (1996) predict electron temperatures of a few 
times $10^7$ to $10^8$ K
in the post--shock flow.
One then might expect to see an emission line spectrum from a corona--like
region above the accretion spot if it could be viewed separately.
The spectra for UZ For, AM Her and VV Pup, extracted at phases when
the spot is just barely obscured behind the limb of the WD
but the EUV emission above the spot is still visible,
are not statistically significantly different in morphology
than the spectra observed when the spot is viewed face on.
There is no compelling evidence for low density discrete plasma line emission.
However, because the signal-to-noise ratio is low in the spectra
from just before and after spot appearence
(due to the fact that the region just above the spot is only visible for
about 0.05 in phase),
the existence of emission lines cannot be entirely precluded.\footnote {
The highly magnetic CV AR UMa, observed at high signal to noise
by EUVE during a high state in December
of 1996, shows no evidence of low density plasma emission lines at any phase,
including those when viewing just above the accretion spot.
Howell \& Sirk, (1998).}
The likely source of the EUV radiation from the column
above the spot is Compton scattering off free electrons 
by the photons emanating from the accretion region below
(cf., ID83 \& below).

UZ For and VV Pup show clear evidence of flux removal 
at phases when the far--field accretion stream crosses the line of sight to the
accretion spot.
Comparing the bright phase spectra (labeled in Figs. 3 and 4) to the 
far--field stream dip spectra shows no
dependence on wavelength during the flux reduction or any sign of
absorption edges.
The slope of the ratio of the far--field 
stream dip spectra to the bright phase spectra
is consistent with zero to within 3 $\sigma$ of the slope uncertainty.
Therefore, we conclude that the stream dips are not primarily 
caused by absorption
from neutral hydrogen (i.e., a cold absorber) but could have components
due to photoelectric absorbtion by 
a warm (ionized) absorber (Cropper et. al., 1997). It is unlikely that
the far-field stream is completely or even mostly ionized (Harrop-Allin
et al., 1997) thus, the number density of free electrons would be
small. We would therefore not expect Compton scattering to play a large
role here.

If the far--field stream near the coupling region was scattering
the EUV photons (Compton scattering at EUV wavelengths causes only about an
8\% change in the initial wavelength), then one might expect to see
observational evidence for this.
If we assume that the coupling region of UZ For 
were a source of EUV radiation it
would be clearly seen immediately following ingress during the first half
of the stellar eclipse when the accretion
region is fully hidden, but with the entire accretion stream still visible.
No EUV flux was detected during the stellar eclipse for the 1995 January UZ For
observation 
(mean eclipse countrate $ = 0.0006 \pm 0.005$ s$^{-1}$, see Fig. 3), however 
the 1993 November light curve shows a very low, but statistically 
significant detection
of flux during eclipse
(mean eclipse countrate $ = 0.005 \pm 0.002$ s$^{-1}$ (WSV)).
During these two observations, the shape and depth of the far--field
absorption dip changed (see Figure 10).
Thus, this may be observational evidence for
the occurrence of far--field scattering of EUV photons.
A more likely explaination, however, is that changes in the far-field
stream position, density, and size cause variations in the optical depth
and photoionization of the stream as a function of phase.

We now consider spectra obtained during the broad dip phases due to
absorption by the column just above the accretion spot.
Comparisons of the spectra extracted from the broad dip and the bright phases
show significant wavelength dependence.
We show spectra extracted during the broad dip phase and the bright phase
for VV Pup, AM Her, and UZ For in the upper panels of Figure 9.
The ratio of the dip phase to the bright phase is shown in the lower panels.
Linear least-squares fits, weighted by the error in the ratio, show
statistically significant positively sloped ratios for all three stars.
The slopes are greater than zero by 12, 32, and 13 $\sigma$ for VV Pup,
AM Her, and UZ For, respectively.
Over the entire wavelength range (75 to 125\AA)
the broad dip phase spectra are {\it softer} than the bright phase spectra.

If the accretion column contained any significant 
neutral hydrogen or helium, then we would expect photoelectric
absorption to occur, producing a typical power-law slope of --3. The slope
seen in Figure 9 is much shallower.
Photoelectric absorption by heavier elements (such as C, N, O, Ne, Mg,
Si, S, \& Fe) is a likely cause of some of the photon loss
in regions close to the shock where
the stream is highly, but not necessarily fully ionized
(Kylafis \& Lamb, 1982). However, no convincing or substantial
absorption lines have been seen in EUV spectra of AM Hers (Craig et al.,
1997), so the
contribution to the observed opacity by ionized metals is small and, at
best, limited to frequencies near absorption edges.
The average photon energies present ($\sim$ 100 \AA\ = 200 eV) are
greater than local blackbody kT values (near 20-40 eV), so Compton
scattering is likely to be the dominant mechanism removing photons.
This scattering process down-converts photon energies and has a
$1-cos\theta$ scattering dependence, so most photons (for 1 scatter)
are scattered back
down towards the WD surface or up towards the accretion column, but with
lower energies in both cases. 
The resulting slope of the spectral distribution will have a direct relation
to the optical depth of the gas if kT is not equal to, or even
approximately near the average photon energy.
The situation here has kT $\sim$ $<E>$.
Since the energy of an EUV photon is much less than the rest energy of
an electron, the Compton scattering cross section is independent of
wavelength.
Thus, flux removal from the line of sight by Compton scattering in 
the accretion stream would be approximately equal at all of our observed
wavelengths, leading to a modest spectral slope.
Thus, we would expect to see softer spectra (as we do
within the broad dips) and some residual EUV emission at right angles to
the column (also as observed, see end of \S 4.2).
The nature of the near--field column is just beginning to be
explored in relation to recent high-energy observations (Cropper et. al.,
1997 and \S 7).

\section{Longterm EUV Variability}

The stars UZ For and RE1149+28 have each been observed three times by
{\it EUVE}, twice during pointed observations with the deep survey/spectrometer,
and once each serendipitously in the scanners by the RAP program.
We display the light curves for all six observations in Figure 10.
For each star, the two faintest light curves have been scaled
to equal the mean countrate (Table 1) of the brightest observation.
In addition to changes in mean EUV flux of up to factors of 3, the shapes
of the light curves and positions in phase of the dip features 
have also changed.
In the eclipsing system UZ For,
the longitude $\psi$ of the accretion spot shifted by over 6\arcdeg\ in just
one month, and the duration in phase of detectable EUV flux increased by 1\%
during the same interval.
The depth and position in phase of both the narrow 
far--field accretion stream dip, and
the broad near--field accretion column dip also changed.
RE1149+28 shows similar, although less pronounced changes.
These clearly varying features indicate dynamic movement of the gas
stream, coupling region, and
accretion regions on or near the WD surface itself.
The eclipsing system HU Aqr has been observed six times over
a 16 month period by {\it EUVE}
specifically to study its longterm behavior  and also has shown a few
degree shift in the spot location. Further details of the longterm
behavior of HU Aqr will be presented in 
Sirk et. al., (1998) \& Schwope et. al., (1998).

\section{Discussion}

\subsection{Correlations between the Accretion Spot Model Parameters} 

Since we have determined some physical sizes and conditions for the EUV
emitting regions in these AM Her stars, it might be useful to examine
any possible correlations which may exist. We remind the reader that
the collection of data used here is used in a mean fashion, thus any
short timescale (less than 1-2 binary orbits) or transient features
of the accretion regions are averaged out. We present results for 
10 different observational data sets on six different stars. The
observations are almost all deep survey (DS) data but two were made with 
the scanner telescopes (ScA, ScB). Our noted correlations should be 
viewed with healthy  skepticism due to the small sample of stars
and intrinsic uncertainties in the physical and model parameters.

Using the information available in Tables 1 \& 2 we find the following
correlations. The ratio of the unabsorbed to absorbed flux, i.e., the
``missing'' flux due to the absorbing near-field column, and the height ($h$) of
the EUV emitting mound are both related to the radius $(r)$ of the accretion
region itself: $Ratio = 4.57r + 0.45$ and $h=0.23r + 0.0076$.
This is not a surprise, as the larger the footprint of the
region, the larger the near-field column and overall emitting height
might be expected to be. The height of the emitting mound 
is also then
correlated with the level of ``missing'' flux. We also find that the
height of the emitting column determined from our model $(h_{col})$
is strongly
related to the white dwarf magnetic field strength: $h_{col} = 0.0026B
+ 0.0094$. 
(This last relation does not include 
the star RE1844
in the fit. This seems reasonable as RE1844 is
known to be a two-pole emitter most of the time; Bailey et al., 1995.)
This $h_{col} - B$ relation is again, not surprising, as
a stronger field should produce a more focused near-field column, leading
to higher temperatures farther above the surface and thus higher
EUV emitting regions.
Comparisons of all other parameters with the magnetic field strength
showed no obvious correlations. 

The values of $r$ and $h$ have some indication of increasing their size with 
an observed increas in the mean EUV count rate.
However, UZ For and RE1149+28
are the only multiple observations we have at hand and not all of these
are with the same EUVE instrument. Thus, a definitive statement is
not possible. We suspect that the overall mean count rate modulates
with the mass accretion rate, but we have no simultaneous data in other
wavebands with which to formulate a firm relationship between these
parameters. Also, we note that from single binary orbital data in the
EUV, small scale changes occur on tens of minute time scales, likely  
indicating regular short-term mass accretion variations.
These aspects of
short-term variability will be discussed by Sirk et al., (1998).

Using contemporaneous EUV and IR observations of HU Aqr (Ciardi, Howell,
\& Hauschildt, 1998) we find similar looking accretion regions (size
and location) but with clear differences in detail. It appears that
changes occur in the accretion regions at all wavelengths but how they
are correlated awaits detailed simultaneous, multi-wavelength observations.

	\subsection{Comparison with Theory} 

Theoretical models of the coupling 
region and accretion spot (Wickramasinghe \& Meggitt; 1985, 
Wickramasinghe; 1989,
Ferrario et. al.; 1989, Beuermann \& Burwitz; 1995, Woelk \& Beuermann;
1996 \& Harrop-Allin et
al.; 1997)
indicate that there should
be a range of magnetic field lines that capture material, the lighter density
regions of the column being grabbed first, the denser blobs later. 
The inhomogeneous stream is composed of regions of high (blobby) 
and low (accretion rain) density blobs.
At the coupling region, the low density material
is striped off first while the larger blobs penetrate further into the
magnetosphere because of their greater ram pressure.
The magnetic field of the WD acts like a giant mass spectrometer.
This separation of big and small blobs is likely to be preserved as
the material proceeds towards the surface of the WD. 

The low density component of the stream forms a shock above the surface
of the WD and X-rays emanate from the post--shock region of the flow.
Roughly one-half of these X-rays are reprocessed
in the WD below and emerge as EUV photons.
The remainder escape freely with some fraction undergoing Compton scattering
in the pre--shock flow (Kylafis \& Lamb, 1982).
The high density (large blob)
part of the stream very likely penetrates the WD surface (King, 1995)
forming a shock that is hidden from view below the surface and
directly heats the surface area producing EUV photons in a region
centered about the dense
part of the stream.
Stockman and Schmidt (1995) suggested that the vertical extent of the accretion
region reported by WSV is caused by Compton scattering within 
the accretion column.
Litchfield (1990) however, modeled blob accretion onto a WD and finds 
that the localized heating distorts the WD photosphere, locally raising it 
by about $\sim 0.4\%$,
sufficient to effect the light curve. The average height of 0.023 R$_{WD}$
we determine for our raised
mound accretion regions  are comparable with those predicted by
Litchfield. 

Thus,
the accretion region on surface of the WD will probably reflect the structure
of the incoming stream leading to an asymmetric, complex footprint.
The assumption would be that 
the low density material would strike nearer the magnetic pole while the
denser material would spread farther away from the pole in magnetic latitude
primarily in the direction of rotation, but also towards the WD equator
(Wickramasinghe et. el., 1991).
Our eclipse data for UZ For and HU Aqr require that the spot is small
($< 0.2$ R$_{WD}$) in rotational longitude.
Brightness gradients in the
longitudinal direction would result in asymmetric 
rise and fall phase light curves if the spot
size is greater than about $0.2$ R$_{WD}$ in longitude.
The light curves for all six self eclipsing systems (Fig. 5)
are symmetric in the rise and fall phase when the accretion region
is scanned vertically
which rules out any large scale, asymmetric structure extened in longitude.

In principle, high time resolution light curves of eclipse ingress and egress
could be used to map out any
longitudinal structures within the accretion regions. However, 
the 2 s duration eclipse profiles in UZ For and HU Aqr
show only a handful of EUV photons which are insufficient for this purpose.
The broad dip features which occur in the light curves 
are asymmetric with respect to the EUV mid--phase, and thus provide the evidence for
small scale ($< 0.2$ R$_{WD}$) geometric and physical 
horizontal structure within the accretion regions.

The high density portion of
the accretion stream is likely to be 
more effective in absorbing or scattering EUV photons than
the low density parts.
We know that the accretion 
stream near the coupling region (i.e., the far--field stream) 
is quite capable of removing EUV flux since we see
the evidence in the narrow stream dips that, at times, 
saturate to zero (in the systems UZ For, EF Eri, HU Aqr, AN UMa, and V834 Cen).
The broad dips, caused by the near--field stream 
very near the WD surface, never remove all the EUV flux and reach zero.
Thus, either the accretion stream near the shock and the WD surface
is very ionized and a much 
less effective photoelectric absorber, or 
the EUV emitting region is larger than the width of the dense part of the
 stream and can never be fully occluded by the stream.
It is likely that both of these effects are present.

Incorporating the ideas disscussed above, we present a scenario that explains
the features observed in the EUV light curves,
the broad dip / bright phase spectral ratios, and the large tilts of the accretion
column required by the column absorber model (\S 4.3).
The schematic diagram in Figure 11 shows an edge--on view of a raised
accretion spot ($h = 0.02$ R$_{WD}$,
mildly extended in WD longitude and latitude with an
accretion column tilted at a 10\arcdeg\ angle away from the magnetic pole
in the direction of rotation.
The high line density indicates the high density portion of the accretion column
where it depresses the otherwise raised accretion region.
Thus, for the proper hydrodynamic conditions (Woelk \& Beuermann,
1996, King 1995) the shock can become buried below the (raised) WD
surface. 
The white portion corresponds to the hotter regions and is the source of
the hardest EUV radiation.
The gray portion illustrates the outer, cooler source of the softer EUV.
The large tilts required in some instances of our absorber modeling are merely a
consequence of using a uniform density accretion column as a first
approximation. A model using a non-uniform accretion column can trade
large tilt angle for higher density regions within the near-field
stream.

The bottom of Figure 11 shows 
a sequence of pictures that illustrate how such a spot appears
when seen rotating across the surface of the WD.
The views at different phases illustrate that the column may 
occlude different regions
of the spot during the binary orbit.
The system inclination and spot parameters in Figure 11 are those of UZ For,
and the three pictures correspond to the rise, broad dip, and bright phases 
marked as such in Figure 3.
Only the dense portion of the accretion column is shown
(as a translucent cylinder).
The inner portion of the spot (white region) that trails the dense 
accretion column is the location of the lower density stream impact
(greater shock height) 
with the WD photosphere.
Again, this inner region is a major source of the harder EUV radiation while
the surrounding outer region is the source of softer EUV radiation.

As the binary rotates, the accretion column will occult 
different parts of the accretion region.
At phases when we look through the dense portion of the stream towards
the the hotter, inner region of the accretion spot,
we expect this inner 
region to be more fully obstructed than the cooler, outer region
resulting in a broad dip phase with a softer spectrum.
This is in fact what we found for our
broad dip / bright phase spectral ratios discussed in \S5..

The non-absorbed geometric model predicts as much as 
twice the radiation than was actually seen during
the central portions of some of the light curves
(see the column labeled ``Ratio'' in Table 2).
What happens to the energy of the radiation that is removed during the various dip
phases?
If the dips are caused solely by Compton scattering, which is at a maximum when we look
parallel to the accretion column, then the radiation simply reappears during the
rise and fall phases when the column presents itself perpendicularly.
If the dips are caused by photoelectric absorption, then the excited material
impacts the WD surface a few seconds later creating an accretion region that is
hotter than it would have been otherwise.

\section{Conclusions}

We have analyzed EUV light curves for ten AM Her systems.  All show strong
modulations as a function of orbital phase. Projection effects account for
the overall bright and faint phases, but cannot explain the sharp transitions
and dips seen in the light curves.
In the systems where the accretion spot rotates behind the limb of the WD 
for some portion of the phase (category 1),
the rise and fall phases of the light curve
are very steep, and are symmetric about the EUV mid-phase. The
observations imply
two things: first, the shape of the steep rise and fall phases is dominated by
vertical extent of the accretion spot, and second, any longitudinal structure
of the accretion region must be less than 0.2 of the R$_{WD}$, since a larger
accretion region would produce an asymmetric lightcurve during the rise and
fall phases.
For the eclipsing systems UZ For and HU Aqr, the short ingress and egress
times directly set upper limits to the spot size of $\le 0.23$ R$_{WD}$.

All the light curves show one or more dips. The eclipsing systems 
constrain the model well and yield
the best results for orbital inclination, spot latitude and longitude,
spot height, and spot size. 
Once the viewing geometry is established, the phasing of the dips restricts
the spatial location of where the accretion stream 
material must reside in order to self-eclipse the accretion region.
The accretion stream, when it is far from the WD surface, quickly crosses the
line of sight to the accretion spot and can completely occlude the spot.
These narrow dips last typically $< 0.1$ of the orbital phase, often
saturate to zero flux, and show no wavelength dependence for the absorption.
The broad dips typically last $\sim 0.25$ of the orbit, are asymmetric in profile,
occur well
before EUV mid phase, never saturate to zero, and show a {\it softer}
spectrum than the non--dip phases.
If the accretion column strikes the accretion region normally, and is concentric
with the spot, the expected light curve would be symmetric and 
show a broad dip at phase 0 as predicted by ID83.
The large asymmetries present 
in all the observed AM Her light curves are {\it direct} evidence
for asymmetries within the accretion regions.
An elongated spot, with the dense portion of the accretion column
eccentrically contacting the spot in the direction of binary 
rotation can explain 
the phasing, duration in phase, and depth of the broad dips.
Furthermore, if the accretion spot shows
a temperature gradient, (hottest towards the center, coolest at the edges), and
the column width is less than the EUV emitting area (or has a
non-uniform density, or both), the broad dips will never
saturate to zero and their spectra will be softer when the dense portion of the
column hides the hottest part of the spot.

Using detailed EUV photometric and spectroscopic observations
of AM Herculis stars, we have constructed an appropriate model of
the accretion regions which include effects of both the far-- and
near--field accretion stream.  
Many of our results below apply equally well to observations of AM
Herculis stars in other high energy bandpasses.
Summarizing our relevant findings we have:

\begin{itemize}
\item [1.]  A flat, circular geometric accretion spot model
only accounts for the gross features of the
EUV light curves, namely the faint phase, the rise to a maximum, and then the
return to the faint phase. This model does not match the
observations during the steep rise and fall phases (which depart from
cosine behavior), for which a vertically extended structure is needed.
We have chosen to model this simply as a hemispherical structure, but
more complex geometries are possible.
 
\item [2.]  All ten stars show large modulations in the form of dips in their
EUV light curves that are inconsistent with cosine projection effects.
All stars where the accretion region is known to lie in the
northern hemisphere (thus guaranteeing that the accretion stream crosses
the line of sight to the accretion region) show a far--field stream eclipse
of the accretion region.
In addition, the two highly inclined systems VV Pup and UZ For, whose
spots are in the southern hemisphere, also show evidence for far--field 
stream eclipses indicating a non--zero width of the coupling region.
Eight stars show additional dip features other than what can be accounted for
by the far--field stream.
The fraction of flux ``missing'' amounts
to as much as 50\% of the flux predicted by the non-absorbed geometric model
(see Table 2).

\item [3.] All AM Herculis stars with multiple EUV observations 
(UZ For, RE1149, QS Tel, HU Aqr), show significant long term variations
in the features of their lightcurves, (eg. the duration of the bright phase,
and the depth and position in phase of both the far--field stream dips, and the
near--field broad dips).
All stars observed also show short term orbit to orbit variations.

\item [4.]  The cause of a majority of the features of the EUV light curves 
is highly dependent
on viewing geometry (i.e., viewing angle to the spot and through the
accretion stream).
The remaining variations are due to the physical structure and
properties of the accretion stream and near--field column themselves. 
Attributing the causes of the features in the light curves 
simply to various astrophysical
effects is incorrect until all the geometrical effects have been taken 
into account and seperated from the physical effects.

\item [5.] The low inclination systems (category 2) 
show more complex EUV light curves and dip features,
attributable again mainly to geometric aspects of 
both the accretion stream and the near--field column. 
The additional complexity observed in these stars 
is probably caused by the fact that the angle subtended between
the spot normal and the viewing direction 
is small and the accretion region is
viewed nearly face on, through a long length of the column, 
for a large fraction of the binary phase
($\sim 25$\%). The accretion region in AM Her is seen nearly face--on
as it crosses the central meridian which (along with its rather weak
magnetic field) may account for it possessing
the most complex lightcurve of the category 1 systems.

\item [6.] Our geometric model matches the EUV observations of the AM Hers very well 
and yields spot sizes averaging 0.06 R$_{WD}$, or
$f \sim 1 \times 10^{-3}$ the WD surface area, and average accretion spot heights
around 0.023 R$_{WD}$.
When the accretion spot undergoes self eclipse (category 1 systems)
the raised mound model is useful for determining the system inclination 
to within 5 degrees,
and the spot colatitude to within 7--15\arcdeg .
The near perfect symmetry of the rise and fall phase of the six systems
shown in Figure 5 preclude any large scale ($> 0.2$ R$_{WD}$) structure
of the accretion region.
The height of the accretion region above the WD surface seems to be
correlated with overall EUV flux (accretion rate), but the spot 
longitude $\psi$ does not (see Table 2).

\item [7.] All category 1 light curves show a small amount of EUV flux from
above the accretion region. This flux can be isolated and observed during the short
phase intervals when the main accretion region itself 
is just hidden from view behind the WD limb.
The EUV spectra during these phases
show no emission lines characteristic of a high temperature,
corona-like plasma, but
signal-to-noise available during these short phases is insufficient to
completely exclude the possibility of weak emission features.
Preliminary analysis of high signal to noise EUV spectra of the highly 
magnetic system AR UMa
(Howell \& Sirk, 1998)
observed during outburst show no evidence of emission lines at any phase.
This result adds weight to the growing body of negative evidence against
the existence of a high temperature, low density plasma (corona)
surrounding the accretion
regions of AM Her stars.

\item [8.] The far--field accretion stream causes narrow dips in the
light curves that saturate to zero in 5 systems: UZ For, AN UMa,
EF Eri, V834 Cen, and HU Aqr, and reduce the flux by a factor of 2 for VV Pup.
The spectra extracted for UZ For and VV Pup during these narrow dip 
phases show no
wavelength dependent variations or absorption features (lines or edges) 
compared with spectra obtained during the out--of--dip phases. 
Thus, in at least these two systems, the far--field accretion stream
behaves like a grey absorber.

\item [9.] A broad dip occurs at binary rotation phases where the far--field
accretion stream is not interfering with our view to the accretion region.
In addition, the broad dips do {\it not} occur at EUV phase zero 
as predicted by assuming
a circular spot impinged normally by a cylindrical accretion column (ID83).
The cause for these broad 
dips must be from material very near the WD surface and appears to be
almost entirely caused by Compton scattering.
The broad dips tell us that either the column is highly inclined
with respect to the WD surface, non-uniform in nature,
or the source of maximum EUV radiation is not
concentric with the densest portion of accretion column.
Since polarimetry data show that the magnetic field is only slightly tilted
with respect to the spot normal in several AM Her stars (Meggitt \& Wickramasinghe, 1989),
we are forced to conclude that the broad dips are caused by small scale
($ < 0.2$ R$_{WD}$) asymmetric structure in both the accretion spot and the
near--field column.

\item [10.] The broad dip / bright phase spectral 
ratio shows a wavelength dependence in that 
the broad dip phase spectra appear softer than the bright phase spectra.
Absorption by a cold absorber (neutral
hydrogen) is not possible due to the high state of, or complete ionization
of H and He and would give a very different spectral slope.
It thus appears that the
broad dips are caused by some (unequal) combination of a warm absorber,
Compton scattering, and geometric effects caused by the near--field stream. 
It is apparent that near the WD surface, the accretion column and region
have a highly asymmetric structure which can significantly 
change on short timescales, days (WSV) to months and longer (see Fig 10,
and Sirk et al., 1998).

\end{itemize}

We have detailed the geometric nature of AM Herculis accretion regions
via the use of high-quality EUV photometric and spectroscopic
observations. Our few--component model provides good fits to the data, but also
indicates some areas where a more complex structure is present.
The magnetic field strength of the WD and the mass accretion rate are likely
to be the dominant mechanisms which cause both the general 
similarities observed
in many of the AM Hers, 
as well as the detailed differences between systems.
The next step in understanding the details presented here 
is to use the presented model,
along with the appropriate physical conditions likely to
be present within the stream and accretion region itself, 
to confirm our results. We need to understand the
roles played by geometry, Compton scattering, photoelectric
absorption, magnetic field strength, mass accretion rate, and other
physical properties, in order to form a better picture of the accretion
regions in AM Hers. 

The authors wish to thank the staff of the Center for Extreme
Ultraviolet Astrophysics for their help throughout the EUVE mission.
SBH and MMS wish to acknowledge partial support of this research by NASA EUVE
grants NAG 5-3523 \& 5-4241 and NASA ADP grants 5-2989 \& 5-3379.
MMS extends thanks to Patrick Sirk for Figure 11.
Adrienne Cool provided many useful comments on an early version of
the manuscript.
Mark Cropper and Axel Schwope have also contributed useful comments.
This manuscript is certified Cruelty Free; no graduate students were
abused in its preperation. Not dishwasher safe.

%
%
%

\clearpage




\makeatletter
\def\jnl@aj{AJ}
\ifx\revtex@jnl\jnl@aj\let\tablebreak=\nl\fi
\makeatother


\begin{deluxetable}{lccrrcc}
\tablewidth{0pt}
\tablecaption{Observation Log}
\tablehead{
\colhead{System } &
\colhead{Instrument$^a$ } &
\colhead{Starting Time } &
\colhead{Duration} &
\colhead{Exposure} &
\colhead{N Orbits$^b$} &
\colhead{Mean Countrate$^c$}\nl
 & & (GMT) & (hours) & (ks) & & (s$^{-1}$) \nl
}
\startdata
UZ For&ScB        & 1993 Oct 16  04:58 & 73 & 98  & 13 & 0.37 \nl
UZ For&DS         & 1993 Nov 18  18:47 & 78 & 85  & 11 & 0.93 \nl
UZ For&DS         & 1995 Jan 15  20:34 & 90 & 82  & 11 & 0.82 \nl
VV Pup&DS         & 1993 Feb 07  21:25 & 34 & 37  & \ 6& 1.53 \nl
AM Her&DS         & 1993 Sep 23  17:57 &114 & 123 & 11 & 2.97 \nl
RE1149+28&DS      & 1993 Feb 22  18:50 & 53 & 64  & 12 & 0.48 \nl
RE1149+28&ScA     & 1994 Mar 08  01:56 &117 & 147 & 27 & 0.26 \nl
RE1149+28&DS      & 1994 Dec 26  06:06 &198 & 145 & 27 & 0.15 \nl
HU Aqr&DS         & 1996 May 29  02:13 & 123& 121 & 16 & 0.044\nl
RE1844-74&DS      & 1994 Aug 17  13:53 & 154& 188 & 35 & 0.65 \nl
EF Eri&DS	  & 1993 Sep 05  13:42 &  99& 107 & 22 & 0.69 \nl
AN UMa&DS         & 1993 Feb 27  22:14 &  27&  33 & \ 5& 0.32 \nl
V834 Cen&DS       & 1993 May 28  03:07 &  35&  37 & \ 6& 0.76 \nl
QS Tel&DS         & 1993 Oct 06  07:51 & 105& 113 & 13 & 1.55 \nl

\tablenotetext{a}{Deep Survey (DS), Scanner A (ScA), and Scanner B (ScB).}
\tablenotetext{b}{Number of full binary orbits sampled.}
\tablenotetext{c}{The Scanner countrates have been multiplied by a factor of 2.2
to account for their smaller effective area compared with that of the 
the Deep Survey in the
Lexan/Boron passband (Sirk et. al., 1997).}

\enddata
\end{deluxetable}

\clearpage




\makeatletter
\def\jnl@aj{AJ}
\ifx\revtex@jnl\jnl@aj\let\tablebreak=\nl\fi
\makeatother


\begin{deluxetable}{lcccccccccccc}
\tablewidth{0pt}
\tablecaption{System Parameter Fits\break
The system inclination is $\iota$.
The angle between the rotational pole and
the EUV accretion spot is $\beta$.
The radius and the height of the accretion spot are $r$ and $h$,
respectively, in units of R$_{WD}$.
The maximum height of the accretion column above the WD surface
that shows significant flux
is $h_{\rm col}$ in units of R$_{WD}$.
The longitude of the accretion spot is $\psi$, the field strength
of the primary magnetic pole is $B$.
The final columns list the ratio of the absorber model flux to the un--absorbed model,
and the absorption coefficient $\tau$ found from the absorber model.
Italic entries for $\iota$, $\beta$, $r$, $h$, and $h_{\rm col}$ are the
parameter solutions derived from our geometric model fits
(roman entries indicate fixed parameters).
Italic entries for values of the period and $\psi$ are 
determined from {\it EUVE} data.
Non-derived entries (roman)
are from Cropper (1990), except where noted.
}
\tablehead{
\colhead{System } &
\colhead{Inst. } &
\colhead{Date } &
\colhead{Period } &
\colhead{$\iota$ } &
\colhead{$\beta$ } &
\colhead{$r$ } &
\colhead{$h$ } &
\colhead{$h_{col}$ } &
\colhead{$\psi$ } &
\colhead{$B$} &
\colhead{Ratio}&
\colhead{$\tau$} \nl
& & & min & ($\arcdeg$) & ($\arcdeg$) & (R$_{WD}$) & (R$_{WD}$) & (R$_{WD}$) & ($\arcdeg$) & (MG)& & \nl
}
\startdata
UZ For&ScB        & Oct 93 & 126.5 &    80.2 & \it114.1 & \it.073 & \it.018 & \it.14 & \it55 & 56 & \it 0.47 & \it 22\nl
UZ For&DS         & Nov 93 & 126.5 & \it80.2 & \it136.6 & \it.060 & \it.031 & \it.15 & \it49 & 56 & \it 0.61 & \it 23\nl
UZ For&DS         & Jan 95 & 126.5 & \it81.7 & \it136.5 & \it.045 & \it.021 & \it.15 & \it49 & 56 & \it 0.61 & \it 23\nl
VV Pup&DS         & Feb 93 & 100.4 & \it73.1 & \it147   & \it.022 & \it.011 & \it.12 &    49 & 32 & \it 0.65 & \it 50\nl
AM Her&DS         & Sep 93 & 185.6 & \it37.1 & \it68    & \it.045 & \it.013 & \it.033&    31 & 12 & \it 0.62 & \it 35\nl
RE1149+28&DS      & Feb 93 & \it90.17& \it70   & \it142   & \it .10 & \it .035& \it.12 & ---   & $43^a$ & \it 0.89 & \it 11 \nl
RE1149+28&ScA     & Mar 94 & \it90.17& \it70  & \it136   & \it.054 & \it .025& \it.12 & ---   & $43^a$  & \it 0.78 & \it 29 \nl
RE1149+28&DS      & Dec 94 & \it90.17& \it70   & \it143   & \it .10 & \it .030& \it.12 & ---   & $43^a$ & \it 0.86 & \it 27 \nl
HU Aqr&DS         & May 96 & 125.0  & \it 81  & \it 40   & \it.036 & \it .021& \it .10& \it 49 & 37$^b$  & --- & --- \nl
RE1844--74&DS     & Aug 94 & 90.10  & \it 73  & \it 144  & \it.090 & \it .025& \it .12& ---   & 10$^c$  & \it 0.94 & \it 8\nl
EF Eri & DS       & Sep 93 & 81.01  & \it 52  & \it 33   & \it.059 &     .025& ---    & ---   & 11$^d$  & ---      & \it 12\nl
\tablenotetext{a}{Schwope, A. D.,1997, private communication}
\tablenotetext{b}{Schwope et. al., (1993)}
\tablenotetext{c}{Ramsay et. al., (1996)}
\tablenotetext{d}{Paerels, F., 1995, private communication}
\enddata
\end{deluxetable}

\clearpage


\makeatletter
\def\jnl@aj{AJ}
\ifx\revtex@jnl\jnl@aj\let\tablebreak=\nl\fi
\makeatother


\begin{deluxetable}{ccccccccc}
\tablewidth{0pt}
\tablecaption{System Geometry Comparison}
\tablehead{
\colhead{System } &
\colhead{Present Work } &
\colhead{} &
\colhead{Polarimetry} &
\colhead{} &
\colhead{Ref.} &
\colhead{Spectrophotometry} &
\colhead{} &
\colhead{Ref.} \nl
 & $\iota$ & $\beta$ & $\iota$ & $\beta$ & & $\iota$ & $\beta$ & \nl
}
\startdata
UZ For & 81 (2) & 136 (5) & 81 (2) & 150 (12) & 1 & 85 (3) & 155 & 2,3 \nl   
HU Aqr & 81 (5) & 40 (10) & 80 (5) & 40 (10)  & 4 & 83 (3) & --- & 5    \nl
VV Pup & 73 (3) & 147 (5) & 74     & 147      & 6 &        &    &         \nl
AM Her & 37 (4) & 68 (8)  & 52     & 49       & 7 &        &    &         \nl
RE1844--74& 73 (5)& 144 (7)& 60    & 167      & 8 &        &    &         \nl
EF Eri & 52 (5) & 33 (10) & 55     & 35       & 6 &        &    &         \nl
\enddata
\tablenotetext{}{Numbers in parentheses are the 1 $\sigma$ errors.
1)From optical eclipse analysis by Bailey \& Cropper, (1991).
2)Beuermann, Thomas \& Schwope, (1988).
3)Schwope, Beuermann \& Thomas, (1990).
4)Glenn et. al., (1994).
5)Schwope, Mantel \& Horne, (1997).
6)Meggitt \& Wickramasinghe, (1989).
7)Wickramasinghe et al., (1991).
8)Bailey et. al., (1995).}
\end{deluxetable}

\clearpage

%
%
%

%


%

\clearpage

\begin{figure}
\caption{Schematic diagram showing AM Her star viewing geometry.
The system inclination is $\iota$, and $\beta$ is the accretion spot colatitude
measured from the rotational pole tilted towards the observer.}

\caption{Geometrical spot model atlas contrasting a flat spot (dashed line)
and raised spot (solid line) for four different inclinations
($\iota = $ 80, 60, 40, and 20\arcdeg),
and a range of spot latitudes that span both hemispheres.
The spot radius is 0.065 R$_{WD}$,
and for the raised mound model,
the height is 0.03 R$_{WD}$ above the WD surface.
All plots are normalized to the maximum brightness when the spot normal
is pointed directly towards the observer.
In all cases, the duration of the bright phase is longer for the raised
mound model. For the higher inclination cases ($\iota = 80\arcdeg$
and $60\arcdeg$)
the rise and fall phase is often steeper for the raised mound model than
for the flat spot.}

\caption{Phase folded mean light curve for UZ For from 1995 January covering
11 full binary orbits binned at 25 s resolution.
Note that the stellar eclipse is total, and the stream
dip also briefly saturates to zero flux. The phases marked ``Rise'' and
``Fall'' are
nearly linear, and very steep in slope. The remaining phases are discussed in
the text.}
\end{figure}

\begin{figure}
\caption{Phase folded mean light curves. The six systems where the accretion
regions undergo self eclipse (category 1) are shown in the first eight panels,
and the remaining four panels show the systems where the accretion region remains
visible at all binary phases (category 2).
Also shown are the geometric model fits
(dotted line), and absorber model fits (solid line).
Open symbols denote data points omitted from the fits.
The model fitting is discussed in the text.
The mean 1 $\sigma$ uncertainty in countrate of all the plotted data points is
indicated at the lower right for each light curve.
The small inset in each plot represents the WD and shows the location of
the accretion
region on the central meridian at phase zero for the denoted inclination
and spot colatitude.}

\caption{Enlargement of the EUV rise and fall phases for the six category 1
AM Her systems shown in Figure 2. The fall phase is mirror
reflected about EUV mid--phase (zero) and shows the near perfect symmetry
of the rise and fall phases when the accretion regions are seen at oblique
angles (edge--on).}

\caption{Model fits to the fall phase of UZ For contrasting the flat spot
(top), raised mound (middle), and raised mound with luminous accretion
column models (bottom).}

\caption{Accretion column brightness as a function of height above WD surface
for the six self eclipsing systems of Figure 4.}
\end{figure}

\begin{figure}
\caption{Schematic geometry of the accretion column absorber model.
The spot radius is $r$, spot height $h$, and the angle the accretion column
makes with the spot normal in the latitude direction is $\alpha$.
For a given viewing direction, the total absorption
is calculated by integrating over all possible path 
lengths (dotted lines) through the accretion column to the spot.}

\caption{Phase-resolved spectra for the bright phase (solid lines),
and broad dip phases (dashed lines) for the
three EUV bright AM Her systems (upper panels). Lower panels plot the ratio
of the broad dip to the bright phase spectra with the thinner lines indicating
the one sigma error in the ratio.
Least squares linear fits to each ratio (plotted as dashed lines)
show slopes significantly greater than zero by
12, 32, and 13 $\sigma$ for VV Pup,
AM Her, and UZ For, respectively,
which clearly indicate softer broad-dip phase spectra.}

\caption{Phase folded EUV light curves for UZ For and RE1149+28 showing long
term changes in accretion morphology. The spot longitude, the near--field
broad dips,
and the far--field 
narrow accretion stream dips all change in depth and phase.
The total duration in phase of detectable flux also changes.
The light curves
are normalized to the brightest observation for each star (solid lines,
see also Table 1).}

\caption{(top panel) Schematic diagram showing an edge view of a raised accretion
spot with accretion column tilted 10\arcdeg~away from the magnetic
pole. The densest portion of the stream, indicated by high line density,
is offset from the center of the spot in the direction away from
the magnetic pole.
(lower panels)
The bottom of this figure shows three views of the raised accretion spot
extended in the direction of WD longitude and latitude
corresponding to the rise phase, broad dip phase, and
EUV maximum phase. The model presented here has used the binary
inclination and accretion spot latitude ($\iota = 81\arcdeg$, $\beta =
135\arcdeg$) of UZ For.
Harder emission (white) arises from
central portion of the accretion spot and is mostly hidden by the column
during the broad dip phase. Softer emission (gray) is visible at all phases.}
\end{figure}

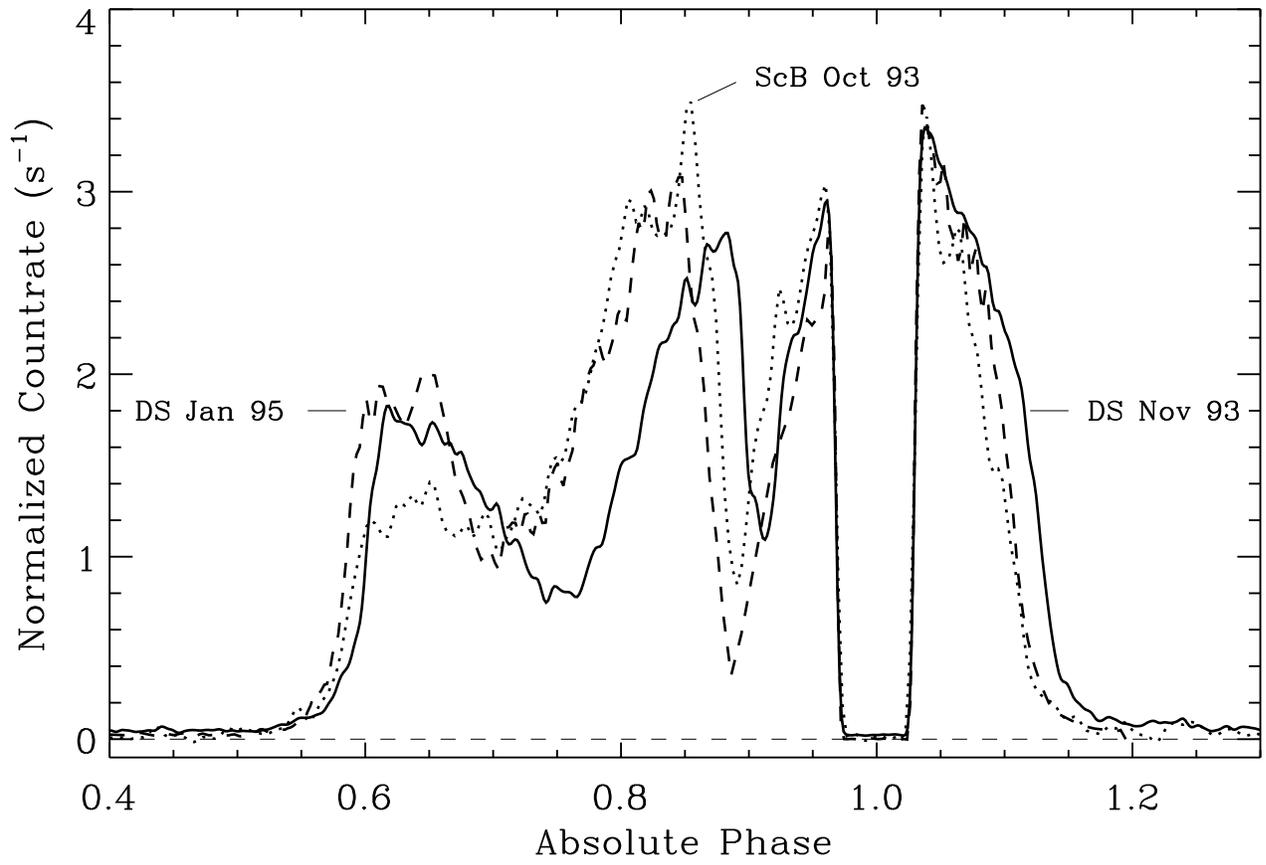
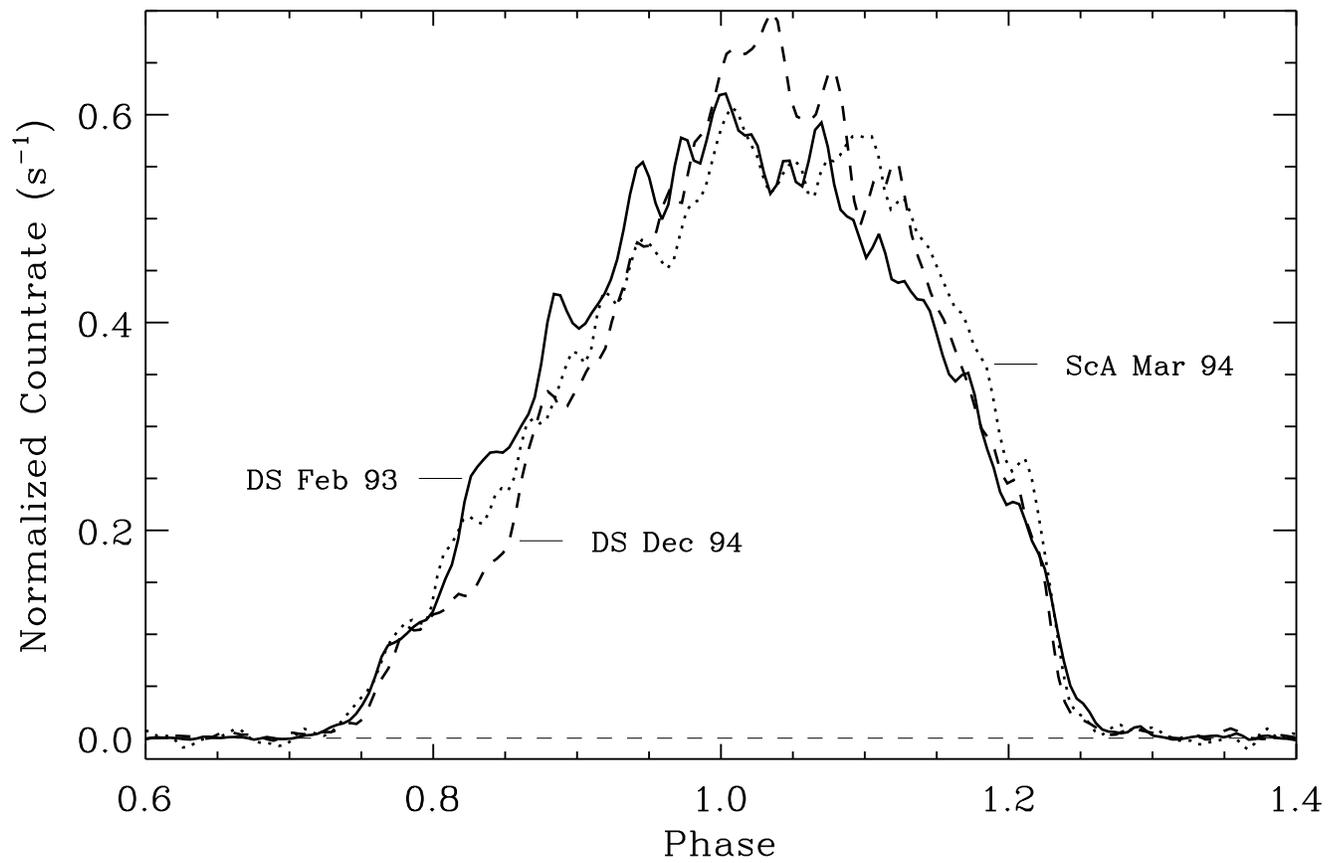
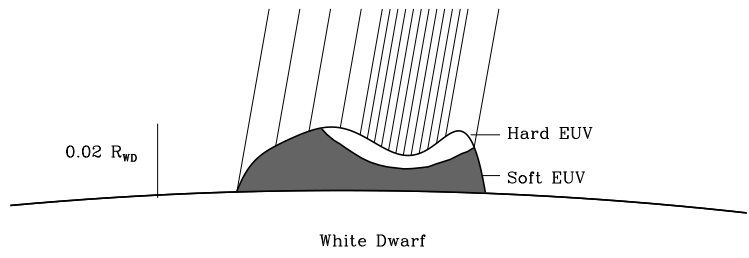
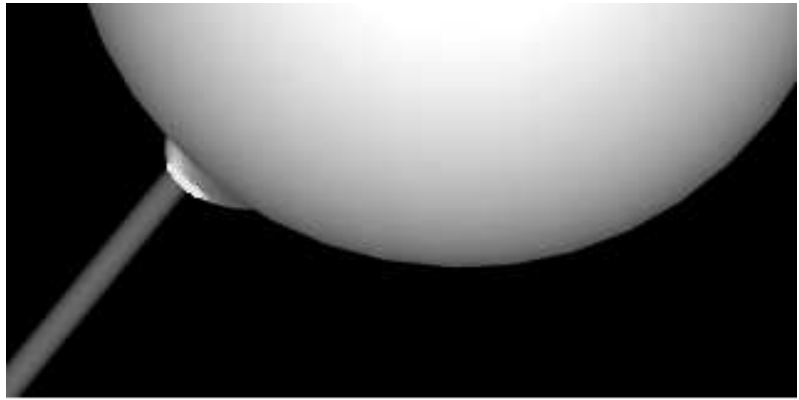
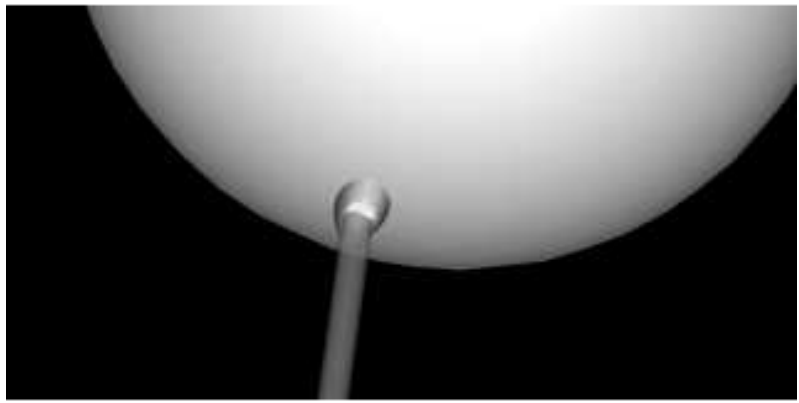
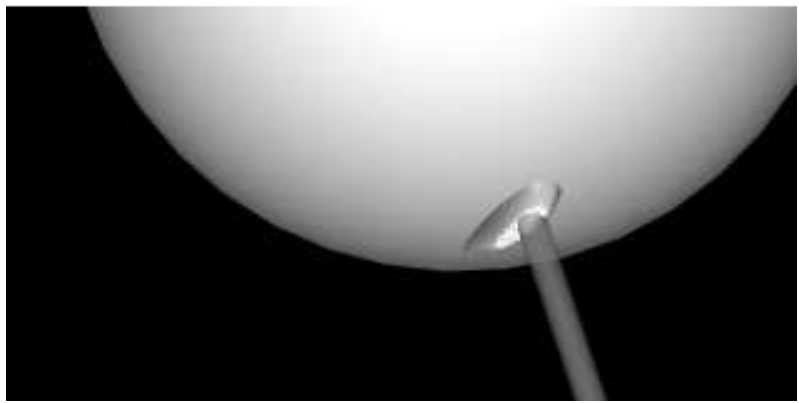
\end{document}